\documentclass[12pt]{article}
\usepackage{amssymb}
\usepackage{amsfonts}
\usepackage{amsmath}
\usepackage[mathscr]{eucal}
\usepackage{amssymb}
\usepackage{amsthm}
\usepackage{mathrsfs}
\usepackage{bm}
\usepackage{graphicx}
\usepackage[english]{babel}
\usepackage[T2A]{fontenc}
\usepackage[utf8]{inputenc}
\usepackage{cite}
\usepackage{ulem}

\newcommand{\be}{\begin{equation}}
\newcommand{\ee}{\end{equation}}
\newcommand{\bea}{\begin{eqnarray}}
\newcommand{\eea}{\end{eqnarray}}

\newcommand{\lb}{\label}
\newcommand{\p}[1]{(\ref{#1})}

\newcounter{rown}

\begin{document}

\begin{titlepage}
\vspace*{0.3cm}

\begin{center}
{\LARGE\bf On BRST Lagrangian description}\\

\vspace{0.5cm}

{\LARGE\bf of partially massless bosonic fields }



%
\vspace{1.5cm}

{\large\bf I.L.\,Buchbinder$^{1,2}$,\,\, S.A.\,Fedoruk$^1$,\,\,  V.A.\,Krykhtin$^{3}$}

\vspace{1.5cm}

\ $^1${\it Bogoliubov Laboratory of Theoretical Physics,
Joint Institute for Nuclear Research, \\
141980 Dubna, Moscow Region, Russia}, \\
{\tt buchbinder@theor.jinr.ru, fedoruk@theor.jinr.ru}

\vskip 0.4cm

\ $^2${\it Department of Theoretical Physics,
Tomsk State Pedagogical University, \\
634041 Tomsk, Russia}

\vskip 0.4cm

\ $^3${\it Tomsk Polytechnic University, 634050 Tomsk, Russia}, \\
{\tt krykhtin@tpu.ru}

\end{center}

\vspace{0.6cm}

\hspace{5.4cm} \verb"Dedicated to the bright memory of I.V. Tyutin"

\vspace{0.6cm}

\nopagebreak

\begin{abstract}
\noindent
We present an exhaustive BRST lagrangian description of partially massless bosonic fields in four-dimensional space.
The basic fields are formulated in terms of two-component spin-tensors in (A)dS space where the tracelessness conditions are automatically fulfilled. The mass shell of partially massless fields is reformulated in terms of constraints on Fock space vectors including the second-class constraints. A conversion procedure for transforming second-class constraints into first-class ones is developed, allowing one to construct a Hermitian and nilpotent BRST charge in the Fock space under consideration. It is proven that the hermiticity and nilpotency restrict the conditions on the theory parameters, which are fulfilled only in dS space. The hermiticity of theBRST charge is incompatible with AdS space. The gauge invariant Lagrangian is constructed on the basis of the BRST charge, and for spin $s$ and depth $t$ the allowed states in the Lagrangian include only $(s-t-1)$ Stückelberg fields. Their exclusion leads to gauge transformations of degree $(s-t)$ for the physical fields. The Lagrangian equations of motion exactly reproduce the mass shell conditions. The Lagrangian in terms of conventional spin-tensor fields is also presented.

\end{abstract}

\vspace{1.0cm}


\phantom{Keywords: }

\newpage

\end{titlepage}

\setcounter{footnote}{0}
\setcounter{equation}{0}

\section{Introduction}
Partially massless fields are a certain type of higher-spin fields.
These fields are defined only in (A)dS space and occupy a kind of
intermediate position between purely massless and purely massive
fields. A specific feature of such fields is that for certain
relationships between the mass parameter and the constant curvature
parameter, the theory is characterized by specific additional gauge
symmetries in comparison with purely massive higher-spin theories.
This yields that partially massless fields, although described by
the same totally symmetric tensors as purely massless or purely
massive ones, nevertheless have a number of corresponding physical
states that lie between the number of states for massive fields and
for massless fields. The term ``partially massless field'' was
introduced in the pioneering papers of Deser and Waldron \cite{DW1,DW2,DW3}\footnote{See also the further developments
in works of the same authors \cite{DW4,DW5,DW6,DW7}.} and Zinoviev \cite{Z}. Some similar aspects of
massive higher spin fields in (A)dS spaces have already been
studies in the earlier papers \cite{DeserN1,DeserN2,Higuchi1,Higuchi2,Higuchi3,Higuchi4,Bengtsson,BMV}, although the term ``partial masslessness'' was not yet used.

The discovery of partially massless fields sparked a surge of
interest in constructing appropriate representations of
the (A)dS group in different dimensions and in developing methods
for describing these fields including Lagrangian formulation,
supersymmetric issues and various applications \cite{DNW,Mets2003,SV,Z1,BIS,S,PV,AG,JLT,JM,BH1,BH2,BBB,
BoH,GHR,BHJR,BKSZ,KZ,KP,BHKP,Mets,L,Bu,HutPon,H,BD,Z2,BLT,Z3,Z4,Z5,Z6,Z7,HJ,BHJJLMM}. We emphasize that
partially massless fields are defined only in (A)dS spaces, and
the aspects of unitarity of the corresponding field theory are highly
nontrivial compared to flat space theories and need to be
specially clarified.

In this paper we focus on developing a general procedure for
constructing a Lagrangian description for an arbitrary theory of partially massless
bosonic fields. The procedure is based on the universal BRST
method for deriving the Lagrangian formulation in a free higher spin
field theory and finding at least cubic vertices as gauge
invariant deformations of a free theory. Lagrangian description for
free massless higher integer spin fields in AdS${}_d$ space was
constructed in \cite{BPT} where the results of such a description
in flat space were used \cite{PT1,PT2,PT3} (see
also a further development in \cite{BBPT,BGK,BKS},
\cite{BFIK} and reference therein). The BRST approach to deriving
the Lagrangian formulation for free massive integer higher spin fields
was developed in \cite{BK} in flat space and in \cite{BKL} in AdS space (see a brief review in \cite{BFK} and references therein). The
Lagrangian formulation for free massless half-integer higher spin
fields was considered in \cite{BKP}.

Application of the BRST approach to higher spin fields
requires a separate discussion of one specific issue. This approach
begins with treating the Lie algebra relations that define
the irreducible field representation of the Poincare or (A)dS group as
first-class constraints in some Fock space. This allows one to
obtain a real gauge invariant Lagrangian in terms of the nilpotent Hermitian BRST
charge while proving that the corresponding equations of motion
reproduce the above-mentioned first-class constraints, ensuring the
correctness of the entire construction. However, in the case of a massive theory, the corresponding
constraints contain ones that form a second-class algebra, and
construction of a nilpotent BRST charge seems impossible. This
difficulty is overcome with the help of the conversion procedure.
This procedure introduces special additional variables that
transform the initial system of second-class constraints into an
equivalent system of first-class ones \cite{FF,FSh,BatF,EM}). In the context of the BRST approach to massive
higher spin theory, the conversion procedure is realized by extending
the Fock space under consideration by an additional set of creation and annihilation operators
(see \cite{PT2,BK,BFK} and references
therein)\footnote{Apparently, the well-known example of conversion is
the reformulation of non-gauge massive electrodynamics (Proca theory) into gauge
invariant theory due to St\"{u}ckelberg's trick (see e.g. \cite{RR}).}.
Note that the BRST approach automatically yields the so-called triplet
Lagrangian formulation (see e.g. \cite{ST}). A more general quartet
formulation of the free massive higher spin fields was given in
\cite{BG}.

In this paper, we consider a derivation of the Lagrangian formulation
for partially massless higher spin fields in terms of totally
symmetric traceless tensors. For simplicity, we restrict ourselves to
the fields in four-dimensional space where the irreducible tensors
$\varphi_{\mu_1\mu_2 \ldots \mu_s}$ can be represented in terms of
two-component dotted and undotted Weyl spin-tensors
$\varphi_{\alpha_1\alpha_2 \ldots
\alpha_s}^{\dot{\alpha}_1\dot{\alpha}_2 \ldots \dot{\alpha}_s}$ (see
e.g. \cite{BuKu}), where the traceless condition is automatically
fulfilled. The BRST approach in higher spin field theory is based on
the relations defining an irreducible representation of the (A)dS group
with arbitrary depth, which are considered as the mass-shell of the
corresponding Lagrangian field theory. In the partially massless
integer higher spin field theory, the mass-shell was described in a
series of previous papers (see, e.g., the recent paper \cite{BH1}
and references therein). We rewrite the mass-shell conditions in
terms of two-component spin-tensors and construct a Hermitian
nilpotent BRST charge using the conversion procedure in the form
considered in \cite{BFK}.

The paper is organized as follows. In Section\,2 we briefly review the conventional relations determining
partially massless irreducible representations in (A)dS${}_4$ space in terms of totally symmetric
traceless fields. All such representations are characterized by two quantities: spin $s$ and depth $t$.
These relations are equivalently rewritten in terms of spin-tensor fields which are totally symmetric
in terms of dotted and undotted indices. In this case, the tracelessness conditions are automatically
fulfilled and all the operations with such fields are simplified. Further we call the relations which describe
the partially massless irreducible representations in terms of fields in (A)dS${}_4$ the mass shell of these
fields.

Section\,3 is devoted to reformulations of the relations determining
the partially massless irreducible representations in
(A)dS${}_4$ space in terms of Fock space vectors. These vectors are
constructed on the basis of spin-tensors introduced in Section 2.
The operators are defined that describe the mass shell of
partially massless higher spin fields in the Fock space. We call
these operators the constraints. A commutator algebra of
constraints is constructed, and it is noted that this algebra
contains both first- and second-class constraints. Gauge
transformations of Fock space vectors are also found in terms of the
Fock space vector parameters.

In Section\,4 we discuss the conversion procedure of the algebra
including the second-class constraints in the theory under
consideration into the algebra with only first-class ones. This is
an important element of the BRST approach to massive higher spin fields
since the BRST charge is defined only for first-class constraint
algebra. In subsection\,4.2 we first  extend the Fock space by new
bosonic creation and annihilation operators that yield extension of
the constraints and second, we extend the Fock space once more
by introducing the fermionic ghost coordinates and momenta which are
ingredients of the general BRST construction. The vectors in the
extended Fock space are introduced and their structure is described.
In subsection\,4.2 we construct the Hermitian and nilpotent BRST
charge and find the converted first-class constraints. It is shown
that the conditions of hermiticity and nilpotency of the BRST charge in
the partially massless theory are fulfilled only in de Sitter
space, AdS space is excluded. In essence, it is independently proved
that unitary partially massless higher spin-theory is well defined
only in de Sitter space.

In Section\,5 we construct the BRST Lagrangian formulation. Gauge
invariant equations of motion are postulated and the corresponding
real gauge invariant Lagrangian is presented. The theory is
described by a large number of fields including the basic field and
the fields related to gauge and auxiliary degrees of freedom. We
show that part of the auxiliary fields can be eliminated
algebraically and the Lagrangian is expressed in terms of the remaining
fields under corresponding gauge transformations. Then, we prove
that the equations of motion and gauge transformations allow one to
eliminate all the auxiliary fields and we arrive at correct fields
and gauge transformations defining the mass shell of the partially
massless higher-spin field theory. This means that the equations of
motion and gauge transformation in terms of the BRST charge reproduce
the conditions defining the partially massless irreducible
representation in the de Sitter space. This completes the BRST
Lagrangian formulation for the fields under consideration. The final
Lagrangian depends only on partially massless fields.

In Section\,6 we present the Lagrangian and gauge transformations in
terms of space-time component fields. Section 7 is a summary
of the results obtained.

\section{Conventional description of partially massless representations in (A)dS${}_4$
space}
In this section we briefly formulate the mass shell field conditions which define the irreducible partially massless
representations on the tensor fields in (A)dS${}_4$ space.

The conventional description of the (A)dS${}_4$ irreducible
representations with a given mass $m$  and a given spin $s$ is
realized on totally symmetric tensor fields $\varphi_{\mu_1 \ldots
\mu_s}=\varphi_{(\mu_1 \ldots \mu_s)}\equiv \varphi_{\mu(s)}$ with
$s$ four-vector indices satisfying the traceless
\be \label{condition0}
g^{\mu_1\mu_2}\varphi_{\mu_1\mu_2 \ldots \mu_s}=0
\ee
and obeying  the conditions (see e.g. \cite{Mets2003})
\be
\label{condition}
\Big\{\Box +\varkappa (s^2-2s-2) -
m^2\Big\}\varphi_{\mu_1 \ldots \mu_s}=0\,, \qquad
\nabla^{\mu_1}\varphi_{\mu_1 \ldots \mu_s}=0\,,
\ee
where $\Box=\nabla^{\mu}\nabla_{\mu}$ is the d'Alambert operator in curved space,
$\nabla_{\mu}$ is the covariant derivative and $\varkappa$ is a
constant in the curvature tensor
\be \label{R-AdS}
R_{\mu\nu}{}^{\lambda\rho}= \varkappa\left(\delta_\mu^\lambda
\delta_\nu^\rho-\delta_\mu^\rho \delta_\nu^\lambda\right).
\ee
In the case of dS space $\varkappa>0$ whereas for AdS space
$\varkappa<0$.

The partially-massless integer higher-spin representations are also
described in terms of the fields $\varphi_{\mu(s)}=\varphi_{(\mu_1 \ldots \mu_s)} $
obeying the conditions \p{condition0} and \p{condition}.
However, in the partially massless case, the mass parameter $m^2$ in \p{condition} is not arbitrary but
is equal to one of the following $s$ values of the quantity $m^2=m_t^2$ where:
\be
\label{m-t}
m_t^2=\varkappa \left(s-t-1\right)\left(s+t\right)
\ee
and the depth parameter $t$ takes one of the following values:
$t=0$ or $t=1$ or ...  or  $t=s-1$.
In this case, equations \p{condition} for a fixed value of the depth parameter $t$ \p{m-t}
are invariant with respect to gauge transformations
\be
\label{tr-vec}
\delta\varphi_{\mu_1 \ldots \mu_s}=
\underbrace{\nabla_{(\mu_1}\ldots\nabla_{\mu_{s-t}}}_{s-t}\lambda_{\mu_{s-t+1} \ldots \mu_s)}\,,
\ee
where the local parameter $\lambda_{\mu_{1} \ldots \mu_t}(x)$ is
the symmetric traceless field
\be
\label{tr-la}
\lambda_{\mu_{1} \ldots \mu_t}=\lambda_{(\mu_{1} \ldots \mu_t)}\,,\qquad
g^{\mu_1\mu_2}\lambda_{\mu_1\mu_2 \ldots \mu_t}=0\,,
\ee
obeying the equations
\be
\label{condition-la}
\Big\{\Box +\varkappa (s^2 +s-t-2) \Big\}\lambda_{\mu_1 \ldots \mu_t}=0\,, \qquad
\nabla^{\mu_1}\lambda_{\mu_1 \ldots \mu_t}=0\,.
\ee
Relations \p{m-t}, \p{tr-vec}, \p{condition-la} were discussed by many authors (see, e.g., \cite{SV,BH1} and references therein). Let us emphasize that equations \p{condition-la} are conditions for
the invariance of equations \p{condition} under gauge transformations \p{tr-vec}.
Note also that expression \p{m-t} implies $m_t^2>0$ in the case of dS space and $m_t^2<0$ for
AdS space.

As we see from \p{m-t}, when the depth parameter is $t=s-1$, the mass parameter is zero: $m_{s-1}=0$.
In this case, we have purely massless fields. Gauge transformations \p{tr-vec} in this case are linear in the covariant derivatives $\nabla_{\mu}$.
For all other values of the depth parameter $t$ ($t=0$ or $t=1$ or ... or  $t=s-2$), we have the fields
describing partially massless representations,
that envolve states with helicities $\pm s,\pm(s-1),\ldots,\pm(t+2),\pm(t+1)$.
In the case of partially massless fields, their gauge transformations \p{tr-vec} are always higher than
first order in the covariant derivatives $\nabla_{\mu}$, in contrast to the purely massless case.
Note that for any fixed $s$, the gauge transformations \p{tr-vec} have maximum degree in derivatives
when the depth parameter is $t=0$. In this case, the parameter  $\lambda_{\mu_{1} \ldots \mu_t}(x)$ is
a scalar $\lambda(x)$ and the variation \p{tr-vec} has the form:\footnote{These systems were considered in \cite{BH1} as some higher spin generalization of
electrodynamics.}
\be
\label{tr-vec0}
\delta\varphi_{\mu(s)}=
\nabla_{\mu(s)}\lambda \,.
\ee

Of course, when none of the conditions \p{m-t} are satisfied (i.e. $m\neq m_t$ for any $t$),
the symmetric traceless field $\varphi_{\mu_1 \ldots \mu_s}$ satisfying equations \p{condition}
does not have any gauge invariance and describes purely massive higher-spin states in (A)dS$_4$ space.

Now we pay attention to that in $4D$ space it is much more convenient to convert each vector index into
a pair of two-component indices $\alpha=1,2$ and $\dot\alpha = \dot{1},\dot{2}$
and deal with spin-tensor fields (defined in the tangent space) of the form\footnote{Note that we consider the covariantly constant frame: $\nabla_\mu e_{m}^{\nu}=0$.}
\be
\label{twocomp}
\varphi^{\dot\beta_{1}\ldots\dot\beta_{s}}_{\alpha_{1}\ldots\alpha_{s}}=
(\sigma^{m_1})_{\alpha_1}{}^{\dot{\beta}_1} \ldots (\sigma^{m_s})_{\alpha_s}{}^{\dot{\beta}_s}
e_{m_1}^{\mu_1}\ldots e_{m_s}^{\mu_s}\varphi_{\mu_1 \ldots \mu_s}
\ee
and their inverses.
If a the spin-tensor field \p{twocomp} is symmetric with respect to its spinor indices
\be
\label{sym-sp-f}
\varphi^{\dot\beta_{1}\ldots\dot\beta_{s}}_{\alpha_{1}\ldots\alpha_{s}}(x)=
\varphi^{(\dot\beta_{1}\ldots\dot\beta_{s})}_{(\alpha_{1}\ldots\alpha_{s})}(x) :=
\varphi^{\dot\beta(s)}_{\alpha(s)}(x)\,,
\ee
the corresponding field with vector indices $\varphi_{\mu_1 \ldots \mu_s}$ is traceless due to the relation
$(\sigma^{m})_{\alpha\dot{\alpha}}(\sigma_{m})_{\beta\dot{\beta}}={}-2\epsilon_{\alpha\beta}\epsilon_{\dot\alpha\dot\beta}$.
That is, for such a field, the tracelessness condition (\ref{condition0}) is already fulfilled automatically.
In what follows, we deal only with traceless fields \p{sym-sp-f}.
In terms of the fields $\varphi^{\dot\beta(s)}_{\alpha(s)}(x)$,
the basic conditions  (\ref{condition}) are rewritten as follows:
\be
\label{condition1}
\Big\{\Box +\varkappa (s^2-2s-2) - m^2\Big\}\varphi^{\dot\beta(s)}_{\alpha(s)}(x)=0\,, \qquad
\nabla^{\alpha_1}_{\dot{\beta_1}}\varphi^{\dot\beta(s)}_{\alpha(s)}(x) =0\,,
\ee
where $\nabla_{\alpha \dot\beta}=(\sigma^{m})_{\alpha\dot\beta} e_{m}^{\mu}\nabla_\mu$.
In this case, the gauge transformations \p{tr-vec} are rewritten in the form
\be
\label{tr-sp}
\delta\varphi_{\alpha_1 \ldots \alpha_s}^{\dot\beta_1 \ldots \dot\beta_s}=
\underbrace{\nabla_{(\alpha_1}^{(\dot\beta_1}\ldots\nabla_{\alpha_{s-t}}^{\dot\beta_{s-t}}}_{s-t}
\lambda_{\alpha_{s-t+1} \ldots \alpha_s)}^{\dot\beta_{s-t+1} \ldots \dot\beta_s)}\,.
\ee
Pay attention to that the  complex conjugation condition for the field \p{sym-sp-f} looks like
\be
\label{sym-sp-f-c}
\bar\varphi^{\beta(s)}_{\dot\alpha(s)}(x)=\left(\varphi^{\dot\beta(s)}_{\alpha(s)}(x)\right)^*,
\ee
when the dotted spinor indices become undotted spinor indices and vice versa.

To conclude this subsection, we pay attention to that in the framework of the formulation in terms of tensor
fields  $\varphi_{\mu(s)}(x)$ or in terms of spin-tensor ones $\varphi^{\dot\beta(s)}_{\alpha(s)}(x)$, derivation of the invariance conditions  \p{tr-vec} or \p{tr-sp},
critical mass values \p{m-t}, and gauge parameter equations \p{condition-la} is technically an extremely
cumbersome procedure. In the next subsection we will show that obtaining these relations is very simplified in the framework of the the Fock space formulation of the theory under consideration.
Note also
that introducing the Fock space formulation is a necessary ingredient for constructing the Lagrangian
BRST formulation.

\setcounter{equation}0
\section{Fock space description of partially massless representations in (A)dS${}_4$
space}

The Fock space  used bellow is constructed on the basis of the spin-tensor fields
$\varphi^{\dot\beta(s)}_{\alpha(s)}(x)$.

We begin with introducing  the bosonic creation
$\bar{c}_{\dot{\alpha}},\,\,c^\alpha$ and annihilation $\bar{a}^{\dot{\alpha}},\,\,a_\alpha$ operators
subject to the commutation relations
\be
[\bar{a}^{\dot{\alpha}},\bar{c}_{\dot{\beta}}]
=\delta^{\dot{\alpha}}_{\dot{\beta}}\,,
\qquad
[a_\alpha,c^\beta]=\delta_\alpha^\beta\,.
\ee
The action of these operators on the vacuum state $|0\rangle$ is defined in the standard way
\be\label{ac-vac}
\langle0|\bar{c}_{\dot{\alpha}}=\langle0|c^\alpha=0\,,
\qquad
\bar{a}^{\dot{\alpha}}|0\rangle=a_\alpha|0\rangle=0
\ee
with the standard conditions for the vacuum
\be\label{vac-cond}
|0\rangle=\left(\langle0|\right)^+\,,\qquad \langle0|0\rangle=1.
\ee
The creation and annihilation operators satisfy the following Hermitian conjugation relations:
 \be
(a_\alpha)^+=\bar{c}_{\dot{\alpha}}\,,\qquad
(\bar{a}^{\dot{\alpha}})^+=c^\alpha\,.
\ee

The Fock space under consideration is a set of  vectors
\be \label{GFState}
|\varphi_{s}\rangle=\frac{1}{s!}\,\varphi^{\dot{\beta}(s)}_{\alpha(s)}(x)\,c^{\alpha(s)}\,
\bar{c}_{\dot{\beta}(s)}|0\rangle
\ee
and conjugate vectors
\be \label{GFState-a}
\langle\bar{\varphi}_s|=\frac{1}{s!}\,\langle 0|\,\bar{a}^{\dot{\alpha}(s)}\,a_{\beta(s)} \bar{\varphi}^{\beta(s)}_{\dot{\alpha}(s)}\,,
\ee
where the field components $\varphi^{\dot{\beta}(s)}_{\alpha(s)}(x)$ and $\bar{\varphi}^{\beta(s)}_{\dot{\alpha}(s)}(x)$
are the fields \p{sym-sp-f-c} and \p{sym-sp-f}, respectively.
Using the commutation relations for the creation and annihilation operators and the explicit form of the vectors
\p{GFState}, it is easy to show that these vectors are eigenvectors for the ``particle number''
operators $K,\, N,\, \bar{N}$:
\be\lb{KNbN}
K=N+\bar{N}+2\,,\qquad N=c^\alpha a_\alpha\,,\qquad
\bar{N}=\bar{c}_{\dot{\alpha}}\bar{a}^{\dot{\alpha}},
\ee
that is:
\be\lb{KNbN-vec}
N|\varphi_{s}\rangle=s|\varphi_{s}\rangle\,,\qquad \bar N|\varphi_{s}\rangle=s|\varphi_{s}\rangle
\,,\qquad
K|\varphi_{s}\rangle=(2s+2)|\varphi_{s}\rangle\,.
\ee

Next we introduce
 the covariant derivative operator
\be\lb{D-t}
D_\mu\ =\ \partial_\mu \ + \ \frac{1}{2}\ \omega_\mu{}^{mn}\,\mathcal{M}_{mn}\,,
\ee
acting on the Fock space vectors where $\omega_\mu{}^{mn}(x)$ is  the spin connection and
\be\lb{M-op}
\mathcal{M}_{mn} \ = \  c^\alpha \,(\sigma_{mn})_\alpha{}^\beta\,a_\beta \ + \
\bar{c}_{\dot{\alpha}}\,(\tilde{\sigma}_{mn})^{\dot{\alpha}}{}_{\dot{\beta}}\,\bar{a}^{\dot{\beta}}
\ee
are the Lorentz algebra generators (see details, e.g., in \cite{BFIK})
\footnote{The spinor conventions, in particular the definition of the matrices $\sigma_{mn}$ and $\tilde{\sigma}_{mn}$, are taken from \cite{BuKu}.}.
Direct calculation of the commutators among the covariant derivative operators \p{D-t} leads to
\begin{equation}
\label{D-mu-alg}
\left[ D_\mu ,D_\nu \right] \ = \ \frac{1}{2}\,R_{\mu\nu}{}^{mn}\mathcal{M}_{mn}
\end{equation}
with the curvature tensor
$R_{\mu\nu}{}^{mn} = R_{\mu\nu}{}^{\lambda\rho} e_{\lambda}{}^m e_{\rho}{}^n$, which has the form \p{R-AdS} in (A)dS spaces.

Now we reformulate  the conditions for the irreducible representation (\ref{condition1}) into the conditions in terms of the Fock space vectors
$\vert\varphi_{s}\rangle$. For this aim, we introduce  operators called constraints:
\begin{eqnarray}
\lb{op-0}
l_0 &=& D^2 \ + \
\varkappa\left(N\bar{N}+N+\bar{N}\right)-m^2\,,
\\ [6pt]
\lb{op-1}
l &=& (a\sigma^m\bar{a})\,e^\mu{}_m D_\mu\,,
\\ [6pt] \lb{op-t1}
l^+&=& -(c\sigma^m\bar{c})\,e^\mu{}_m D_\mu\,,
\end{eqnarray}
where
\be\lb{D2}
D^2 \
=\ g^{\mu\nu}\left(D_\mu D_\nu-\Gamma_{\mu\nu}^\lambda
D_\lambda\right) \ =\ \frac{1}{\sqrt{-g}}\, D_\mu
\sqrt{-g}g^{\mu\nu}D_\nu
\ee
and, as usual, $g=\det g_{\mu\nu}$,
$g_{\mu\nu}=e_{\mu}{}^m e_{\nu}{}_m$.
Using the above relations, one can prove that
\be\lb{D2-N2}
D^2 |\varphi_{s}\rangle
\ =\ \nabla^2|\varphi_{s}\rangle \ =\ \Box|\varphi_{s}\rangle\,,
\ee
where $\Box=\nabla^{\mu}\nabla_{\mu}$ and $\nabla_{\mu}$ are the conventional covariant derivatives
which act on the component fields $\varphi_{\mu(s)}(x) \sim \varphi^{\dot\beta(s)}_{\alpha(s)}(x)$
of the vector $|\varphi_{s}\rangle$ in the standard way.

It is easy to check that the operators \p{op-0}, (\ref{op-1}) and \p{op-t1} form an algebra in terms
of commutators:
\be\lb{com-ll-6}
\left[l^+,l \right] \ = \ K\,(l_0+m^2)\,,
\ee
\be\label{algebra-r}
[l,l_0]=2\varkappa\,(K+1)\,l\,,\qquad [l_0,l^+]=2\varkappa\,(K-1)\,l^+\,.
\ee
As we see, this algebra is closed only in the purely massless case $m=0$.
However, at $m\neq 0$, in the language of quantization of a constrained system, this is a typical algebra for systems with second-class constraints. Direct calculations also
show that only nonzero commutators of the operators \p{op-0}, (\ref{op-1}) and \p{op-t1}
with the operators (\ref{KNbN}) have the form:
\be\label{N-l}
[{l},N]=[{l},\bar{N}]=[{l},K/2]={l}
\,,\qquad [{l}^+,N]=[{l}^+,\bar{N}]=[{l}^+,K/2]={}-{l}^+\,.
\ee

We are now ready to present the on-shell equations  (\ref{condition1}) and gauge transformations \p{tr-sp}
in terms of the Fock space vectors $\vert{\varphi}_{s}\rangle$.

First, acting by the operators ${l}_0$, $l$ defined in \p{op-0}, \p{op-1}
on the vectors $|\varphi_{s}\rangle$, one obtains that
equations  (\ref{condition1}) are rewritten as follows:
\be
\label{condition-vect}
\Big\{ {l}_0-2\varkappa (2s+1) \Big\} |\varphi_{s}\rangle=0\,,\qquad
{l}|\varphi_{s}\rangle=0\,.
\ee

Second, the gauge transformations \p{tr-sp} on the vectors
$|\varphi_{s}\rangle$ take the form
\be
\label{tr-vect}
\delta|\varphi_{s}\rangle= (l^+)^{s-t} |\lambda_{t}\rangle,
\ee
where the vector $|\lambda_{t}\rangle$ describing the  gauge transformation parameters has
the form similar to vector \p{GFState}
\be
\label{la-State}
|\lambda_{t}\rangle=\frac{1}{t!}\,
\lambda^{\dot{\beta}(t)}_{\alpha(t)}(x)\,c^{\alpha(t)}\,\bar{c}_{\dot{\beta}(t)}|0\rangle\,.
\ee

Note that the system of basic equations \p{condition-vect} must be preserved
under the transformations \p{tr-vect}. This requirement imposes conditions on the vector
of gauge transformation parameters $|\lambda_{t}\rangle$ defined in \p{la-State}.
An explicit realization of these conditions is simply obtained if we carry out the analysis in terms
of the Fock space vectors.
So using the relations
\be\lb{l0-l}
l_0(l^+)^n=(l^+)^n\big[l_0+2\varkappa n(K+n)\big]\,,
\ee
\be\lb{l-ln}
l(l^+)^n={}-n\bigr(K-n+1\bigl)(l^+)^{n-1}
\Bigr[l_0+m^2+\varkappa (n-1)(K+n)\Bigl]\ +\ (l^+)^{n}l
\ee
and taking into account the commutators
\be
K(l^+)^n =(l^+)^n\big(K+2n\big)\,,\qquad
K|\lambda_{t}\rangle=2(t+1)|\lambda_{t}\rangle\,,
\ee
we obtain that in order to preserve the conditions \p{condition-vect}
under gauge transformations \p{tr-vect},
the vector of gauge transformation parameters must obey the following conditions:
\be
\label{condition-vect-la}
\Big\{ {l}_0+2\varkappa (s-t-1)(s+t+1)\Big\} |\lambda_{t}\rangle=0\,,\qquad
{l}|\lambda_{t}\rangle=0\,.
\ee
In this case, for a given $t$, the mass parameter $m$ represented in ${l}_0$
must take the value $m=m_t$, where $m_t$ exactly coincides with the value
given in \p{m-t}.
As a result, we see that relations \p{condition-vect-la} for $|\lambda_{t}\rangle$ exactly
coincide with conditions \p{condition-la} for the gauge parameters
$\lambda^{\dot{\beta}(t)}_{\alpha(t)}(x)\sim \lambda_{\mu(t)}(x)$.

\setcounter{equation}{0} \setcounter{equation}{0}
\section{Conversion of second-class constraints into first-class ones
and construction of BRST charge}

Within the BRST approach to free higher spin field theories, the
gauge invariant Lagrangian is constructed on the basis of the Hermitian
and nilpotent BRST charge\footnote{To be more precise, here we mean
the canonical BRST charge or the BRST-BFV charge \cite{FV},
\cite{BV}, \cite{FF} (see also \cite{HT}). Initially, the BRST-BFV approach
was developed for quantization of gauge theories. The application of this approach to Lagrangian formulation of
higher-spin field theories
is an illustration of utility and power of quantum methods in classical field theory.}.
At the same time, the Hermiticity ensures the reality of the Lagrangian
while the nilpotency ensures gauge invariance. It is important, the BRST
charge is constructed on the basis of relations defining the mass-shell of the theory under consideration. In the context of the BRST
approach, it is natural to call the above relations  constraints.
The procedure for constructing a nilpotent BRST charge assumes that
the constraint algebra is closed. In the language of canonical
quantization of constraint systems, this means that nilpotency of
the BRST charge requires a set of first-class constraints and its
hermiticity assumes that the algebra is invariant under Hermitian
conjugation. In the case of partially massless fields under
consideration, the relations defining the mass-shell is
formulated in terms of the operators $l_0$ \p{op-0}, $l$ \p{op-1},
$l^{\dagger}$ \p{op-t1} playing the role of constraints. Commutator
algebra of these operators  is given by relations \p{com-ll-6},
\p{algebra-r}. We see that at $m \neq 0$, this algebra is indeed
invariant under Hermitian conjugation; however, it is not closed. In
the language of canonical quantization of constrained systems, we
have a case with second-class constraints, and construction of a
BRST charge looks like impossible. This obstacle is avoided with
the help of  a conversion procedure (see e.g., \cite{FF,FSh,BatF,EM}).

The conversion procedure is based on extending the second-class
constraints by additional variables so that the new constraints
form the first-class constraint system (see e.g., \cite{FF,FSh,BatF,EM}). In the context of the BRST approach to
massive higher spin field theories, the constraint procedure assumes
introducing new creation and annihilation operators and extending the initial Fock space
(see, e.g., discussion in \cite{BFK} and references therein). As a
result, we arrive at a theory with a greater number of auxiliary fields
than in the initial theory but nevertheless the equations of motion
in the theory after conversion should reproduce the mass-shell of
the initial theory. In the case of partially massless theory under
consideration, we introduce the new bosonic creation and annihilation
operators $b^+$ and $b$, extend the Fock space constructed in
the previous section by these new operators and add additional terms containing
$b^+$ and $b$ to the constraints
\p{op-0}, \p{op-1},  \p{op-t1}. This leads to a set of new first-class constraints that
allows one to construct the Hermitian and nilpotent BRST charge in terms
of new constraints in the extended Fock space.

\subsection{Extensions of the Fock space}

Further consideration requires two extensions of the initial Fock
space spanned by the operators $c$ and $\bar{c}$.

Commutation relations \p{com-ll-6} show that the constraints
$l$ and $l^+$ are second-class ones. To convert these
constraints into first-class ones, we add to the oscillators
$\bar{c}_{\dot{\alpha}}$, $c^\alpha$ and $\bar{a}^{\dot{\alpha}}$
$a_\alpha$, the new bosonic oscillators $b$ and $b^+=(b)^{+}$ subject to
the standard bosonic commutation relations:
\be
[b,b^+]=1\,.
\label{b-com}
\ee
Then, according to the conversion procedure, we
should add special terms with the operators $b$ and $b^+=(b)^{+}$ to
the constraints $l_0$, $l$, $l^{\dagger}$ to get a first-class
commutator algebra. This automatically means that we should construct
a new Fock space based not only on the operators
$\bar{c}_{\dot{\alpha}}$, $c^\alpha$ and  $\bar{a}^{\dot{\alpha}}$,
$a_\alpha$ but also on the operators $b$ and $b^+$. Then, we should
construct the BRST charge using the new first-class constraints.

Let us move on to constructing the BRST charge in the converted
theory. We begin with the Fock space spanned by the oscillators
$\bar{c}_{\dot{\alpha}}$, $c^\alpha$, $\bar{a}^{\dot{\alpha}}$,
$a_\alpha$, $b$, $b^+$ and following the BFV prescription \cite{FV},
\cite{BV}, \cite{FF}, extend this Fock space by the fermionic ghost
coordinate operators  $\eta_{0}$, $\eta$, ${\eta}^{+}$ and the
corresponding ghost momenta $\mathcal{P}_{0}$, $\mathcal{P}^{+}$,
$\mathcal{P}$. The ghosts $\eta_{0}$ and $\mathcal{P}_{0}$ are
Hermitian, $\eta^ {+}_{0}=\eta_{0}$,
$\mathcal{P}^{+}_{0}=\mathcal{P}_{0}$, whereas
$(\eta)^{+}={\eta}^{+}$, $(\mathcal{P})^{+}=\mathcal{P}^{+}$. The
only nonzero anticommutators for the ghost variables have the form
\be \{\eta,\mathcal{P}^+\}
 \ = \ \{\mathcal{P}, \eta^+\}
 \ = \ \{\eta_0,\mathcal{P}_0\}
 \ = \ 1.
\label{ghosts} \ee The ghost operators are characterized by the
following ghost numbers: $gh(\eta_0)= gh(\eta)=gh(\eta^+)=1$,
$gh(\mathcal{P}_0)=gh(\mathcal{P})=gh(\mathcal{P}^+)=-1$. The ghost
numbers for all bosonic operators are zero.

The extended vacuum $|0\rangle$ is defined, besides \p{ac-vac}, by
the conditions \be b|0\rangle  \ = \ 0\,,\qquad \langle 0|\,b^+  \ =
\ 0\,, \label{b-vac} \ee \be \lb{ghost-vac} \eta|0\rangle \ = \
\mathcal{P}|0\rangle \ = \ \mathcal{P}_0|0\rangle \ = \ 0 \,,\qquad
\langle 0|\,\mathcal{P}^+  \ = \ \langle 0|\,\eta^+  \ = \ \langle
0|\,\mathcal{P}_0  \ = \ 0\,. \ee

By definition, the vectors of the extended BRST space under
consideration should have zero ghost number and, therefore, the most
general form for these vectors looks like
\be \label{Phi-s}
|\Psi_s\rangle= |\Phi_s\rangle \ + \
\eta_0\mathcal{P}^+|\mathcal{Z}_{s-1}\rangle \ + \
\eta^+\mathcal{P}^+|\mathcal{X}_{s-2}\rangle \,.
\ee
The Fock space
vectors $|\Phi_s\rangle$, $|\mathcal{Z}_{s-1}\rangle$ and
$|\mathcal{X}_{s-2}\rangle$ are expanded in powers of the oscillators
$b^+$, where the coefficients of this expansion are the Fock space
vectors spanned by the oscillators $c$ and $\bar{c}$. As a result,
we get the vectors $|\Phi_s\rangle$, $|\mathcal{Z}_{s-1}\rangle$ and
$|\mathcal{X}_{s-2}\rangle$ in the form
\begin{eqnarray}
\label{GFState-g} &&\displaystyle |\Phi_{s}\rangle= \sum_{k=0}^{s}
(b^+)^{s-k} |\varphi_{k}\rangle\,,\qquad\qquad |\varphi_{k}\rangle=
\frac{1}{k!}\,\varphi^{\dot{\beta}(k)}_{\alpha(k)}(x)\,c^{\alpha(k)}\,
\bar{c}_{\dot{\beta}(k)}|0\rangle\,,\qquad\\
\label{GFState-g-1} &&\displaystyle |\mathcal{Z}_{s-1}\rangle=
\sum_{k=0}^{s-1} (b^+)^{s-1-k} |\zeta_{k}\rangle\,, \qquad \
|\zeta_{k}\rangle=
\frac{1}{k!}\,\zeta{}^{\dot{\beta}(k)}_{\alpha(k)}(x)\,c^{\alpha(k)}\,
\bar{c}_{\dot{\beta}(k)}|0\rangle\,,
\\
\label{GFState-g-2} &&\displaystyle |\mathcal{X}_{s-2}\rangle=
\sum_{k=0}^{s-2} (b^+)^{s-2-k} |\chi_{k}\rangle\,, \qquad
|\chi_{k}\rangle=
\frac{1}{k!}\,\chi{}^{\dot{\beta}(k)}_{\alpha(k)}(x)\,c^{\alpha(k)}\,
\bar{c}_{\dot{\beta}(k)}|0\rangle\,.
\end{eqnarray}
All the vectors $|\varphi_{k}\rangle$ and $|\zeta_{k}\rangle$,
$|\chi_{k}\rangle$ have the form similar to \p{GFState}. In
relation \p{GFState-g}, we restricted the expansion so that it does
not contain states with spin higher than $s$. Here the vector
$|\varphi_{s}\rangle$ contains the basic field with a given spin $s$, all
other vectors $|\varphi_{k}\rangle$ with $k < s$ are
auxiliary fields. The vectors $|\mathcal{Z}_{s-1}\rangle$ and
$|\mathcal{X}_{s-2}\rangle$ contain only auxiliary fields with spins
less than $s$.

The conjugate extended vector
\be \label{bPhi-s} \langle\bar\Psi_s|=
\langle\bar\Phi_s| \ + \
\langle\bar{\mathcal{Z}}_{s-1}|\mathcal{P}\eta_0 \ + \
\langle\bar{\mathcal{X}}_{s-2}|\mathcal{P}\eta \,,
\ee
contains the
component vectors:
\begin{eqnarray}\label{GFState-g-a}
&&\displaystyle \langle\bar{\Phi}_{s}|= \sum_{k=0}^{s}
\langle\bar{\varphi}_{k}| (b)^{s-k}\,, \qquad\qquad
\langle\bar{\varphi}_k|=\frac{1}{k!}\,\langle 0|\,\bar{a}^{\dot{\alpha}(k)}\,a_{\beta(k)} \bar{\varphi}^{\beta(k)}_{\dot{\alpha}(k)}\,,\\
\label{GFState-g-a-1} &&\displaystyle
\langle\bar{\mathcal{Z}}_{s-1}|= \sum_{k=0}^{s-1}
\langle\bar{\zeta}_{k}| (b)^{s-1-k}\,, \qquad\
\langle\bar{\zeta}_{k}|=\frac{1}{k!}\,\langle
0|\,\bar{a}^{\dot{\alpha}(k)}\,a_{\beta(k)}
\bar{\zeta}{}^{\beta(k)}_{\dot{\alpha}(k)}\ \,,
\\
\label{GFState-g-a-2} &&\displaystyle
\langle\bar{\mathcal{X}}_{s-2}|= \sum_{k=0}^{s-2}
\langle\bar{\chi}_{k}| (b)^{s-2-k}\,, \qquad\
\langle\bar{\chi}_{k}|=\frac{1}{k!}\,\langle
0|\,\bar{a}^{\dot{\alpha}(k)}\,a_{\beta(k)}
\bar{\chi}{}^{\beta(k)}_{\dot{\alpha}(k)}\,.
\end{eqnarray}
As a result, we arrive at a triplet of vectors $|\Phi_{s}\rangle$,
$|\mathcal{Z}_{s-1}\rangle$, $|\mathcal{X}_{s-2}\rangle$ and their
conjugate vectors $\langle\bar{\Phi}_{s}|$,
$\langle\bar{\mathcal{Z}}_{s-1}|$, $\langle\bar{\mathcal{X}}_{s-2}|$
in the Fock space corresponding to the oscillators
$\bar{c}_{\dot{\alpha}}$, $c^\alpha$, $\bar{a}^{\dot{\alpha}}$,
$a_\alpha$, $b$, $b^+$.

Let us introduce the operator $S$ by the rule \be \label{K-form}
S=\frac{1}{2}\,\big(N+\bar N\big) +\mathrm{n} =
\frac{1}{2}(K-2)+\mathrm{n}\,,
\ee
where the operator $\mathrm{n}$ has
the form
\be \label{n-op}
\mathrm{n}:=b^+ b\,.
\ee
Then it is easy to show that the vectors
\p{GFState-g}--\p{GFState-g-2}, \p{GFState-g-a}--\p{GFState-g-a-2} are the eigenvectors for the
operator $S$: \be\label{K-phi} S|\Phi_{s}\rangle=s|\Phi_{s}\rangle
\,,\qquad
S|{\mathcal{Z}}_{s-1}\rangle=(s-1)|{\mathcal{Z}}_{s-1}\rangle
\,,\qquad
S|{\mathcal{X}}_{s-2}\rangle=(s-2)|{\mathcal{X}}_{s-2}\rangle \,,
\ee \be\label{K-phi-b}
\langle\bar{\Phi}_{s}|S=s\langle\bar{\Phi}_{s}| \,,\qquad
\langle\bar{\mathcal{Z}}_{s-1}|S=(s-1)\langle\bar{\mathcal{Z}}_{s-1}|
\,,\qquad
\langle\bar{\mathcal{X}}_{s-2}|S=(s-2)\langle{\mathcal{X}}_{s-2}|
\,.
\ee
This allows  treating the operator $S$ as a spin operator in
the Fock space corresponding to the oscillators
$\bar{c}_{\dot{\alpha}}$, $c^\alpha$, $\bar{a}^{\dot{\alpha}}$,
$a_\alpha$, $b$, $b^+$. Now we define the Hermitian operator
$\mathcal{S}$  as follows:
\be
\label{cS-def} \mathcal{S}=S
+\eta^+\mathcal{P}+\mathcal{P}^+\eta \,.
\ee
Then one can see that
the extended space vectors \p{Phi-s} and \p{bPhi-s} are the
eigenvectors of this operator:
\be\label{S-phi}
\mathcal{S}|\Psi_{s}\rangle=s|\Psi_{s}\rangle \,,\qquad
\langle\bar{\Psi}_{s}|\mathcal{S}=s\langle\bar{\Psi}_{s}| \,.
\ee
Therefore, the operator $\mathcal{S}$ is the spin operator in
the extended Fock space with the vectors $|\Psi_{s}\rangle$ \p{Phi-s}.

\subsection{Construction of the converted constraints and the Hermitian and nilpotent BRST charge}

In fact, up to this point we have not used the constraint details.
Therefore, the above discussion should be valid for any massive
higher spin theory. Now we will  explicitly construct the
deformation of the initial constraints \p{op-0}, \p{op-1}, \p{op-t1}
due to the operators $b$, $b^+$ and construct the Hermitian and
nilpotent BRST charge based on these constraints. The conditions of the above hermiticity and nilpotency
will prove to truncate the form of the vectors \p{GFState-g}, \p{GFState-g-1},
\p{GFState-g-2}, \p{GFState-g-a}, \p{GFState-g-a-1},
\p{GFState-g-a-2} and single out only de Sitter space. This means
that the partially massive higher spin theory is well defined only
in de Sitter space.

In the case under consideration, the conversion procedure consists in
a suitable modification of the  constraints ${l}$, ${l}^+$, and
${l}_0$ by additional $b$\,- and $b^+$- dependent terms.
Therefore, we study the following type of converted
constraints:
\be \label{el}
{L}_0 = l_0+A_0(\mathrm{n})\,,\qquad {L}
= l+A(\mathrm{n})b\,,\qquad {L}^{+}= l^{+}+b^{+}
{A}^{+}(\mathrm{n})\,.
\ee
where ${l}_0$, ${l}$, ${l}^+$ are the
initial constraints \p{op-0}, \p{op-1}, \p{op-t1}. Additional second
parts $A_0(\mathrm{n})$, $A(\mathrm{n})b$, $b^+ {A}^{+}(\mathrm{n})$
in the converted constraints ${L}_0$, ${L}$, ${L}^+$
depend,  respectively, on the annihilation and creation operators $b$ and $b^+$,
where $A_0(\mathrm{n})$, $A(\mathrm{n})$ and $A^+(\mathrm{n})$ are some unknown yet functions of the
operator $\mathrm{n}$ defined in \p{n-op}.
Note that the
spin operator $\mathcal{S}$ defined in \p{cS-def} preserves the
operators \p{el} defined in \p{cS-def}:
\be\label{K-L}
\big[{L}_0,\mathcal{S}\big]=0\,,\qquad
\big[{L},\mathcal{S}\big]={L}\,,\qquad
\big[{L}^+,\mathcal{S}\big]={}-{L}^+\,.
\ee

The explicit forms of the functions $A_0(\mathrm{n})$ and $A(\mathrm{n})$ will be determined by the
requirement that the BRST operator built on the basis of the new constraints ${L}$, ${L}$, ${L}^{+}$ is
Hermitian and nilpotent on the space of the states \p{Phi-s} and \p{bPhi-s}.
That is, we will construct such operators ${L}$, ${L}$, ${L}^{+}$ and
such the BRST charge $Q$ that the following conditions hold:
\be\label{Q-0}
\langle\bar{\Psi}_{s}|Q^+|\Psi_{s}\rangle=\langle\bar{\Psi}_{s}|Q|\Psi_{s}\rangle\,,\qquad
\langle\bar{\Psi}_{s}|Q^2|\Psi_{s}\rangle=0\,.
\ee
The first is the condition of hermiticity and the second one is the condition of nilpotency in the extended
Fock space of the vectors $|\Psi_{s}\rangle$ \p{Phi-s}. The second of relations \p{Q-0} be evidently
fulfilled if the operator  $Q^2$ has the form
\be\label{Q-2}
Q^2=\mathcal{B}\left(\mathcal{S}-s \right)+\left(\mathcal{S}-s \right)\mathcal{B}^+\,,
\ee
where $\mathcal{B}$ is some arbitrary operator. Bellow we will construct the BRST charge satisfying the
relation \p{Q-2}.

To construct the BRST charge in explicit form we use the analogy with massless higher integer spin theory
in (A)dS space. First of all, we proceed with that the converted constraints should form a first-class
constraint algebra in (A)dS space. Such an algebra was considered in \cite{BFIK}. Therefore, we suppose
that the converted constraints $L_0$, $L$, $L^+$ satisfy the same constraint algebra in
(A)dS space as the first-class constraints for the massless theory in
(A)dS space. Thus, by analogy with the constraint algebra from \cite{BFIK}, we consider the following
algebra for the converted constraints
$\big[{L}^+,{L}\big]= {K} {L}_0$,
$\big[{L},{L}_0\big]=2\varkappa\,( {K}+1)\,L$,
$\big[{L}_0,{L}^+\big]=2\varkappa\,{L}^+\,( {K}+1)$,
which should be fulfilled in the space of the vectors $|\Psi_{s}\rangle$ \p{Phi-s}. Taking into account
this algebra, we immediately get the Hermitian BRST charge in the form
\begin{eqnarray}
Q&=&
\eta_0{L}_0+\eta^+{L}+\eta\, {L}^++ {K}\eta^+\eta\mathcal{P}_0
\nonumber
\\[0.5em]
&&{}
-2\varkappa( {K}-1)\eta_0\eta \mathcal{P}^+
+2\varkappa( {K}+1)\eta_0\eta^+\mathcal{P}
-4\varkappa\, \eta_0\eta^+\eta\,\mathcal{P}^+\mathcal{P}
\,,
\label{QAdS}
\end{eqnarray}
which is similar to the BRST charge for massless higher-spin fields in AdS${}_4$ space,
considered in \cite{BFIK}.

The operator $Q^2$ is obtained from expression \p{QAdS} by direct calculations with help of
relations \p{com-ll-6}, \p{algebra-r}, \p{N-l}. It has the form
\begin{eqnarray}
Q^2&=&
\eta\eta^+\Big\{\big[{L}^+,{L}\big]- {K} {L}_0 \Big\}
\nonumber
\\[0.5em]
&&{}
+\eta_0\eta\Big\{[{L}_0,{L}^+]-2\varkappa\,( {K}-1)\,{L}^+
\ -\ 4\varkappa\, \eta^+\mathcal{P}\,b^+{A}^{+}(\mathrm{n}) \Big\}
\nonumber
\\[0.5em]
&&{}
{} +
\eta^+\eta_0 \Big\{[{L},{L}_0]-2\varkappa\,( {K}+1)\,{L}
\ +\ 4\varkappa\, \eta \mathcal{P}^+\,A(\mathrm{n})b \Big\}
\,.
\label{QAdS2a}
\end{eqnarray}
Then to obtain \p{Q-2}, we must postulate the following commutation relations:
\be
\label{THEequations2a}
\big[{L}^+,{L}\big]= {K} {L}_0+\alpha\big(S+1-s\big)\,,
\ee
\be\label{THEequations2b}
\big[{L},{L}_0\big]=2\varkappa\,( {K}+1)\,{L}+\beta\big(S+1-s\big)\,,\qquad
\big[{L}_0,{L}^+\big]=2\varkappa\,{L}^+\,( {K}+1) +\big(S+1-s\big)\beta^+\,,
\ee
where $\alpha$ is some operator  independent of any ghosts (it can be treated as a constant) and
\be
\label{betta}
\beta={}-4\varkappa\,A(\mathrm{n})b\,,\qquad
\beta^+={}-4\varkappa\,b^+{A}^{+}(\mathrm{n})\,.
\ee
With this choice of algebra, expression \p{QAdS2a} takes the form
\be
Q^2\ =\ \alpha\,\eta\eta^+ (\mathcal{S}-s) \ +\ \beta\,\eta^+\eta_0 (\mathcal{S}-s)
\ +\ (\mathcal{S}-s) \eta_0\eta \beta^+\,,
\label{com-el}
\ee
Thus, the condition \p{Q-2} is satisfied and hence the operator \p{QAdS} is nilpotent on average
\p{Q-0} in the space of the vectors \p{Phi-s}, \p{bPhi-s}.

All that remains is to find the functions $A_0(\mathrm{n})$ and $A(\mathrm{n})$ and the constant
$\alpha$ from the requirement the commutation relations \p{THEequations2a} and \p{THEequations2b}
are fulfilled.

So using the relations
\be
b A(\mathrm{n})=A(\mathrm{n+1})b\,,\qquad
b^+ A(\mathrm{n})=A(\mathrm{n-1})b^+
\label{A-b-com}
\ee
and similar for $A_0$, ${A}^+$ (these relations based on \p{b-com}), one gets
\be
\label{LL-com-a}
\big[b^+ {A}^+(\mathrm{n}),A(\mathrm{n})b\big]=
{}\mathrm{n}A(\mathrm{n}-1){A}^+(\mathrm{n}-1)-
(\mathrm{n}+1)A(\mathrm{n})A^+(\mathrm{n})\,.
\ee
Inserting the operators \p{el} in \p{THEequations2a}
and using \p{com-ll-6} and \p{LL-com-a},
we obtain the condition
\be
\label{cond-a}
K\Big[m^2-A_0(n)\Big]
+\mathrm{n}A(\mathrm{n}-1){A}^+(\mathrm{n}-1)-
(\mathrm{n}+1)A(\mathrm{n}){A}^+(\mathrm{n}) \ = \
\alpha\big(S+1-s\big)\,.
\ee
In addition, substituting the operators \p{el} in the first equality of \p{THEequations2b},
and using \p{algebra-r} and \p{A-b-com},
we obtain the condition
\be
\label{cond-b}
A(\mathrm{n})b\Big[A_0(\mathrm{n})-A_0(\mathrm{n}-1)-
2\varkappa(K+1)\Big] \ = \
\beta\big(S+1-s\big)\,.
\ee
After using the expression for $\beta$ from
\p{betta}, equation
\p{cond-b} leads to the equality
\be
A_0(\mathrm{n})-A_0(\mathrm{n}-1)=2\varkappa (2s-2\mathrm{n}+1)\,,
\ee
which has the solution
\be\lb{A0}
A_0(\mathrm{n})=2\varkappa\,\mathrm{n} \big(2s-\mathrm{n} \big)\,.
\ee
Note that an additional dimensional constant term can also be added to expression \p{A0}.
However, such an additional term is undesirable in the purely massless case $m_t=0$, where the depth is
$t=s-1$ and no conversion procedure is required. For this reason, such an additional term is not present in expression \p{A0}.

After finding $A_0(\mathrm{n})$, we now find $A(\mathrm{n})$.
To do this, we substitute expressions \p{K-form} and \p{A0}
into equation \p{cond-a}. As a result, we get
\be
\label{cond-a1}
K\Big[m^2-2\varkappa\,\mathrm{n} \big(2s-\mathrm{n} \big)\Big]
+\mathrm{n}A(\mathrm{n}-1){A}^+(\mathrm{n}-1)-
(\mathrm{n}+1)A(\mathrm{n}){A}^+(\mathrm{n}) \ = \
{\displaystyle\frac12}\,\alpha K+\alpha\big(\mathrm{n}-s\big)\,,
\ee
which leads to the condition
\be
\alpha=2\Big[m^2-2\varkappa\,\mathrm{n} \big(2s-\mathrm{n} \big)\Big]
\ee
as well as the equation
\be
\label{cond-a2}
\mathrm{n}A(\mathrm{n}-1){A}^+(\mathrm{n}-1)-
(\mathrm{n}+1)A(\mathrm{n}){A}^+(\mathrm{n}) \ = \ 2\Big[m^2-2\varkappa\,\mathrm{n} \big(2s-\mathrm{n} \big)\Big]\big(\mathrm{n}-s\big)\,.
\ee
This equation \p{cond-a2} has the following solution:
\be
\label{cond-As2}
A(\mathrm{n}){A}^+(\mathrm{n}) \ = \ \big(2s-\mathrm{n} \big)
\Big[m^2-\varkappa\,\mathrm{n} \big(2s-\mathrm{n}-1 \big)\Big]\,.
\ee
Note that in expression \p{cond-As2}, similarly to expression \eqref{A0} for the operator $A_0(\mathrm{n})$, the correct limit for the purely massless case $m_t=0$ is taken into account.

Finally, from  \p{cond-As2} we get
\be
\label{sol-As}
\begin{array}{rcl}
A(\mathrm{n}) &=& \varepsilon\,\big(2s-\mathrm{n} \big)^{1/2}
\Big|m^2-\varkappa\,\mathrm{n} \big(2s-\mathrm{n}-1 \big)\Big|^{1/2}\,,
\\ [6pt]
{A}^+(\mathrm{n})  &=& \bar\varepsilon \big(2s-\mathrm{n} \big)^{1/2}
\Big|m^2-\varkappa\,\mathrm{n} \big(2s-\mathrm{n}-1 \big)\Big|^{1/2}\,,
\end{array}
\ee
where $|\varepsilon|=1$. Without loss of generality, we can set $\varepsilon=1$
and
\be
\label{sol-As-f}
A(\mathrm{n}) \ = \
{A}^+(\mathrm{n}) \ = \ \big(2s-\mathrm{n} \big)^{1/2}
\Big|m^2-\varkappa\,\mathrm{n} \big(2s-\mathrm{n}-1 \big)\Big|^{1/2}\,,
\ee

As a result, we obtain the operators ${L}_0$, ${L}$ and ${L}^+$ in \p{el} in which
$A_0(\mathrm{n})$, $A(\mathrm{n})$ and ${A}^+(\mathrm{n})$ have the forms \p{A0} and \p{sol-As-f}.
The converted constraints are finally found. Direct check shows that for given $A(\mathrm{n})$ and ${A}^+(\mathrm{n})$ the needed relations \p{THEequations2a} and \p{THEequations2b} are indeed fulfilled.

Now let us turn to relation \p{cond-As2}. In essence, this relation is a direct
consequence of the Hermitian and nilpotent nature of the BRST charge. As we will see, it yields the
extremely important conclusions.

The left-hand side of relation \p{cond-As2}
is the Hermitian operator. Therefore, the operator on the right-hand side
\be\lb{op-AA}
\big(2s-\mathrm{n} \big)
\Big[m^2-\varkappa\,\mathrm{n} \big(2s-\mathrm{n}-1 \big)\Big]
\ee
must take non-negative values on the vectors under consideration.
On the vectors
\be\lb{vect-exp}
|\phi_{k}\rangle=
(b^+)^{s-k}
|\varphi_{k}\rangle\,,
\ee
where $k$ takes possible values from $k=0$ to $k=s$,
that form the vector \p{GFState-g}, the operator \p{op-AA} takes the value
\be\lb{const-AA}
\big(s+k\big)
\Big[m^2-\varkappa\,\big(s-k\big) \big(s+k-1 \big)\Big]\,,
\ee
since $\mathrm{n}(b^+)^{s-k}|0\rangle = (s-k)(b^+)^{s-k}|0\rangle$.
However, in the case of partially massless representations under consideration, $m^2=m_t^2$, where $m_t^2$ is determined by expression \p{m-t}.
Substituting this value instead of $m^2$ in expression \p{const-AA}, we obtain the condition for the
parameters of the theory
\be\lb{cond-pm}
\varkappa \Big[(s-t-1)(s+t)\,- \,\big(s-k\big) \big(s+k-1 \big)\Big]\geq 0\,.
\ee
This condition, on the one hand, is a consequence of hermiticity and nilpotency of the BRST charge, but on
the another hand, it is a condition of unitarity of the partially massless theory. Therefore, one concludes
that the unitarity condition is closely related to the basic properties of the BRST charge.

The condition \p{cond-pm} effectively leads to a change in the range of summation in the expansions
\p{GFState-g},  \p{GFState-g-1}  \p{GFState-g-2},
restricting the parameter $k$ in these sums to smaller numbers.

Consider the Sitter space with $\varkappa>0$. Then the condition \p{cond-pm} for fixed values of $s$ and $t<s$  is
satisfied for the following values of $k$ in expansion  \p{GFState-g} for $|\Phi_{s}\rangle$:
\be\lb{k-pm-dS}
k=t+1,t+2,\ldots,s-1,s \,.
\ee
Thus, in this case the expansion  \p{GFState-g} has the form:
\be\label{GFState-pm}
|\Phi_{s}\rangle=
\sum_{k=t+1}^{s} (b^+)^{s-k}
|\varphi_{k}\rangle\,.
\ee
Analogously, expansions \p{GFState-g-1} and \p{GFState-g-2} for the fields $|\mathcal{Z}_{s-1}\rangle$
and $|\mathcal{X}_{s-2}\rangle$, respectively, take the forms:
\be\label{GFState-pm-1}
|\mathcal{Z}_{s-1}\rangle=
\sum_{k=t}^{s-1} (b^+)^{s-1-k}
|\zeta_{k}\rangle\,, \qquad
|\mathcal{X}_{s-2}\rangle=
\sum_{k=t-1}^{s-2} (b^+)^{s-2-k}
|\chi_{k}\rangle\,.
\ee
Note that at $t=0$ the summation over $k$ in the expansion of $|\mathcal{X}_{s-2}\rangle$ starts with the value $k=0$.
Pay attention to that the depth $t$- dependent truncation in the above expansion is important
for obtaining a correct degree of derivatives in gauge transformations of partially massless fields.

In AdS space, where $\varkappa<0$,  the condition \p{cond-pm}
is not satisfied at least for vectors with a given spin $s$, which must necessarily be present in
the partially massless representation containing helicities $\pm s,\pm(s-1),\ldots,\pm(t+2),\pm(t+1)$.
This means that in the case of AdS space, the hermiticity and nilpotency conditions of the BRST
are violated and the Lagrangian formulation for partially massless higher-spin fields is impossible
in the framework of the BRST approach. In essence,
the fundamental properties of the BRST charge uniquely distinguish the Sitter space as a true vacuum for
partially massless higher spin fields.

To conclude this section, let us emphasize some properties of the operator $A(\mathrm{n})$ at $m^2=m^2_t$
\p{sol-As-f} in dS space. It is easy to see that the vectors $(b^+)^k|\varphi_{l}\rangle$ at any $l$ are the
eigenvectors of this operator with eigenvalues
\be\label{Ak-k}
A(k)= \varkappa^{1/2}(2s-k)^{1/2}\Big|(s-t-1)(s+t)-k(2s-k-1)\Big|^{1/2}\,.
\ee
The direct consequences of this expression are the identity
\be\label{Ak0}
A(s+t)=0\,,\qquad A(s-t-1)=0\,,
\ee
which will be used bellow.

\setcounter{equation}{0}
\section{Lagrangian description of partially massless fields}

\subsection{BRST Lagrangian and equations of motion}

The master equation in the BRST approach to higher spin field theory is postulated in terms of the BRST charge in the form:
\begin{eqnarray}\label{eqQ}
Q\,|\Psi_s\rangle \ = \ 0\,,
\end{eqnarray}
where the BRST charge $Q$ is defined in \p{QAdS}.
Due to the nilpotency of the BRST charge $Q$, the vector $|\Psi_s\rangle$
in  equation (\ref{eqQ}) is defined up to gauge transformations
\begin{eqnarray}
|\Psi'_s\rangle \ = \ |\Psi_s\rangle \ + \ Q\,|\bm{\Lambda}_s\rangle \,,
\label{gauge transf}
\end{eqnarray}
where $|\bm{\Lambda}_s\rangle$ is the extended Fock space valued gauge parameter
with the ghost number $-1$. Taking into account that the ghost number of BRST charge is $1$ and the ghost number of the vector $|\Psi_s\rangle$ is $0$, we can present the gauge parameter in the form
\begin{eqnarray}\label{gtQ}
|\bm{\Lambda}_s\rangle \ = \ \mathcal{P}^+|\Lambda_{s-1}\rangle\,,
\label{gauge parameter}
\end{eqnarray}
where the gauge parameter $|\Lambda_{s-1}\rangle$ in (\ref{gauge parameter}) has now the ghost number $0$ and
is given the decomposition in the extended Fock space
\be\label{gauge-par-ex-pm}
|\Lambda_{s-1}\rangle =
\sum_{k=t}^{s-1} (b^+)^{s-1-k}
|\lambda_{k}\rangle\,,\qquad
|\lambda_{k}\rangle=\frac{1}{k!}\,\lambda^{\dot{\beta}(k)}_{\alpha(k)}(x)\,c^{\alpha(k)}\,\bar{c}_{\dot{\beta}(k)}|0\rangle\,.
\ee
analogous to $|\mathcal{Z}_{s-1}\rangle$ in \eqref{GFState-pm-1}.

The Lagrangian leading to the equation of motion \p{eqQ} is written in the form
\be
{\cal L}_{s,t}
\ = \
\int d\eta_0\; \langle\bar\Psi_s|\,Q\,|\Psi_s\rangle
\,.
\label{actionQ}
\ee
Due to the nilpotency of the BRST charge, the Lagrangian \p{actionQ} is gauge invariant, as are the equations of
motion \p{eqQ}.

The equations of motion \p{eqQ} are written for the vector of the extended Fock space depending on ghost variables.
To clarify and simplify the structure of the equations
of motion, we rewrite the equations of motion in terms of the vectors $|\Phi_{s}\rangle$, $|{\mathcal{Z}}_{s-1}\rangle$ and
$|{\mathcal{X}}_{s-2}\rangle$, which do not depend on ghost variables and are the components of the vector
$|\Psi_s\rangle$ in expansion \p{Phi-s}. For this aim, we
integrate the Lagrangian \p{actionQ} over ghost variables and arrive at the Lagrangian in terms of a triplet
of component ghost-independent vectors in the form
\begin{eqnarray}
\mathcal{L}_s
&=&
\langle\bar\Phi_s|\Bigl\{
\Bigl[{L}_0-2\varkappa(K-1)\Bigr]|\Phi_s\rangle-{L}^+|{\mathcal{Z}}_{s-1}\rangle
\Bigr\}
\nonumber
\\
&&{}
-\langle\bar{\mathcal{Z}}_{s-1}|\Bigl\{
{L}\,|\Phi_s\rangle-{L}^+|{\mathcal{X}}_{s-2}\rangle+K|{\mathcal{Z}}_{s-1}\rangle
\Bigr\}
\nonumber
\\
&&{}
-\langle\bar{\mathcal{X}}_{s-2}|\Bigl\{
\Bigl[{L}_0+2\varkappa(K+1)\Bigr]|{\mathcal{X}}_{s-2}\rangle-{L}\,|{\mathcal{Z}}_{s-1}\rangle
\Bigr\}.
\label{LagrAdS}
\end{eqnarray}

The equations of motion for the vector triplet $|\Phi_{s}\rangle$, $|{\mathcal{Z}}_{s-1}\rangle$ and
$|{\mathcal{X}}_{s-2}\rangle$ follow from the Lagrangian \p{LagrAdS} in the form:
\begin{eqnarray}
&&
\Bigl[{L}_0-2\varkappa(K-1)\Bigr]|\Phi_s\rangle-{L}^+|{\mathcal{Z}}_{s-1}\rangle
=0\,,
\label{236}
\\[5pt]
&&{}
{L}\,|\Phi_s\rangle-{L}^+|{\mathcal{X}}_{s-2}\rangle+K|{\mathcal{Z}}_{s-1}\rangle
=0\,,
\label{237}
\\ [5pt]
&&{}
\Bigl[{L}_0+2\varkappa(K+1)\Bigr]|{\mathcal{X}}_{s-2}\rangle-{L}\,|{\mathcal{Z}}_{s-1}\rangle
=0.
\label{238}
\end{eqnarray}
The gauge transformation \p{gauge transf} can also be rewritten as a system of gauge transformations for the above triplet
\be
\delta|\Phi_s\rangle \ = \ {L}^+\,|\Lambda_{s-1}\rangle\,,
\qquad
\delta|{\mathcal{X}}_{s-2}\rangle \ = \ {L}\,|\Lambda_{s-1}\rangle\,,
\label{GTAdS0}
\ee
\be
\delta|{\mathcal{Z}}_{s-1}\rangle \ = \ {L}_0\,|\Lambda_{s-1}\rangle\,.
\label{GTAdS0a}
\ee
Further, we examine this system of equations and gauge transformations in more detail.

First of all, we note that the vector $|{\mathcal{Z}}_{s-1}\rangle$
describing the states $|\zeta_{t}\rangle, |\zeta_{t+1}\rangle, \cdots , |\zeta_{s-2}\rangle,
|\zeta_{s-1}\rangle$ is purely auxiliary and can be eliminated algebraically with the help of the equation of motion \p{237}:
\be
|{\mathcal{Z}}_{s-1}\rangle={}-K^{-1}\Big({L}\,|\Phi_s\rangle-{L}^+|{\mathcal{X}}_{s-2}\rangle\Big).
\label{1-solut}
\ee
Here we take into account that the operator $K$ (\ref{KNbN}) is non-singular and the inverse operator $K^{-1}$ is well defined. In particular,
for an arbitrary vector $|\psi_n\rangle$
the following relation holds:
\be
K^{-1}\,|\psi_n\rangle=\frac{1}{2n+2}\,|\psi_n\rangle\,.
\ee
Substituting the result (\ref{1-solut}) into Lagrangian (\ref{LagrAdS}), one gets the Lagrangian in terms of the vectors $|\Phi_s\rangle$ and $|{\mathcal{X}}_{s-2}\rangle$ in the form
\begin{eqnarray}
\nonumber
\mathcal{L}_{s,t}
&=&
\langle\bar\Phi_s|\Big({L}_0 -2\varkappa(K-1) + {L}^+ K^{-1} {L}\Big)|\Phi_s\rangle
\\ [5pt]
&&
{}- \langle\bar{\mathcal{X}}_{s-2}|\Big({L}_0 +2\varkappa(K+1) - {L} K^{-1} {L}^+\Big)|{\mathcal{X}}_{s-2}\rangle
\label{Lagr-vector-1}
\\ [5pt]
&&{}
-\langle\bar{\mathcal{X}}_{s-2}|\,{L} K^{-1} {L}\,|\Phi_s\rangle
- \langle\bar\Phi_s|\,{L}^+ K^{-1} {L}^+\,|{\mathcal{X}}_{s-2}\rangle \,,
\nonumber
\end{eqnarray}
The corresponding equations of motion are written as follows:
\begin{eqnarray}
&&
\Bigl({L}_0 -2\varkappa(K-1) + {L}^+ K^{-1} {L}\Bigr)|\Phi_s\rangle \ - \ {L}^+ K^{-1} {L}^+\,|{\mathcal{X}}_{s-2}\rangle \
= \ 0 \,,
\label{eqs-s}
\\ [5pt]
&&{}
\Bigl({L}_0 +2\varkappa(K+1) - {L} K^{-1} {L}^+\Bigr)|{\mathcal{X}}_{s-2}\rangle \ + \
{L} K^{-1} {L}\,|\Phi_{s}\rangle
\ =\ 0\,.
\label{eqs-2s}
\end{eqnarray}
Both Lagrangian (\ref{Lagr-vector-1}) and the equations of motion
(\ref{eqs-s}) and (\ref{eqs-2s}) are still gauge invariant under the transformations \p{GTAdS0} for the
vectors $|\Phi_s\rangle$ and $|{\mathcal{X}}_{s-2}\rangle$.

Gauge invariant Lagrangian \p{Lagr-vector-1} is the partially massless counterparts of the Zinoviev
Lagrangian for the massive field with the St\"{u}ckelberg gauge symmetry \cite{Zino}.
Specifically, the vector $|\Phi_{s}\rangle$ contains states
$|\varphi_{t+1}\rangle,  \cdots ,
|\varphi_{s-1}\rangle$ and $|\varphi_{s}\rangle$, among which the states
$|\varphi_{t+1}\rangle,  \cdots ,
|\varphi_{s-1}\rangle$ are Zinoviev-like gauge ones (St\"{u}ckelberg states),
and the vector $|\varphi_{s}\rangle$ describes physical helicities. All other vectors correspond to gauge and auxiliary degrees of freedom.

Note that purely massless fields in (A)dS space were considered in a similar approach in \cite{BFIK}.
In our consideration, it corresponds to the case when the depth parameter is equal to $t=s-1$.
Then in \p{GFState-g}, \p{GFState-g-1}, \p{GFState-g-2}
there is only one state in each extended vector: $|\Phi_{s}\rangle=|\varphi_{s}\rangle$,
$|{\mathcal{Z}}_{s-1}\rangle=|\zeta_{s-1}\rangle$, $|{\mathcal{X}}_{s-2}\rangle=|\chi_{s-2}\rangle$.

Relations \p{eqQ}, \p{gauge transf}, \p{actionQ} in essence solve the problem of constructing a Lagrangian
formulation for partially massless higher-spin theory in dS space. To fully complete the construction, it is
necessary to answer the remaining open question regarding the connection of these relations with the original
mass-shell conditions \p{condition-vect}, \p{tr-vect}. The matter is that in the framework of the BRST approach
we firstly extended the original Fock space introduced in Section\,3 by additional creation and annihilation
operators $b^+$, $b$ and ghosts and secondly, we postulated the equation of motion \p{eqQ}. As a result, the
vectors $|\Psi_s\rangle$ contain a large number of gauge and auxiliary degrees of freedom, which are absent on
the mass-shell. For the final validity of the approach under consideration, it is necessary to prove that the
equations of motion \p{eqQ} and gauge transformations \p{gauge transf} identically reproduce the mass-shell
conditions \p{condition-vect}, \p{tr-vect}. This will be done in the next subsection.

\subsection{Derivation of mass-shell conditions}

In this subsection we will show that the equations of motion \p{eqQ} and gauge transformations
\p{gauge transf} do indeed eliminate all the auxiliary degrees of freedom and lead to the correct mass shell
in the theory under consideration. The procedure of deriving the mass shell is in principle similar to one in
the purely massless theory considered in \cite{BFIK}. However, as we shall see, in the partially massless case
there is an important difference that arises at the gauge fixing stage. Namely, after partial gauge fixing,
we obtain for physical fields the correct remaining gauge transformations of higher-order in derivatives
\p{tr-vect} (or the same \p{tr-sp}) than in the purely massless theory. Moreover, the order $(s-t)$ of derivatives
in the gauge transformations coincides with the number of auxiliary Stückelberg gauge fields. The further
analysis is carried out within the framework of the Fronstal type Lagrangian \p{Lagr-vector-1}, which is
obtained after eliminating from the original Lagrangian \p{LagrAdS} the auxiliary BRST fields
$|\zeta_{t}\rangle, \cdots , |\zeta_{s-1}\rangle$ contained in the vector $|{\mathcal{Z}}_{s-1}\rangle$ of
the extended Fock space.

In the purely massless case, partial eliminating of the gauge degrees of freedom is achieved by imposing the suitable gauges (see e.g. \cite{BFIK} and reference therein). \footnote{These gauge fixing conditions can be called the generalized de Donder conditions by analogy with the purely massless theory (see e.g. \cite{Ponomarev} and reference therein.)} We will use its generalization at the first step of our consideration here. In the case under consideration, we
consider the following gauge condition:
\be
{L}\,|\Phi_s\rangle-{L}^+|{\mathcal{X}}_{s-2}\rangle=0
\label{dD-g-vec}
\ee
in terms of the BRST-vectors $|\Phi_s\rangle$ and $|{\mathcal{X}}_{s-2}\rangle$. First of all, one notes that this condition can indeed be imposed. It means, if the vector ${L}\,|\Phi_s\rangle-{L}^+|{\mathcal{X}}_{s-2}\rangle = |f\rangle$ is nonzero, there exists a suitable gauge vector parameter that $|f\rangle \neq 0$ can be done vanished after gauge transformation to the new fields $|\Phi_s\rangle'$ and $|{\mathcal{X}}_{s-2}\rangle'$. Gauge transformation of the above relation yields the equation ${L}\,|\Phi_s\rangle'-{L}^+|{\mathcal{X}}_{s-2}\rangle' +[L,L^+]|\Lambda_{s-1}\rangle = |f\rangle$. Taking into account the commutator \p{THEequations2a} and that $S|\Lambda_{s-1}\rangle = (s-1)|\Lambda_{s-1}\rangle$, one gets the equation for $|\Lambda_{s-1}\rangle$ in the form
$-KL_0|\Lambda_{s-1}\rangle = |f\rangle$. Since the operator $KL_0$ is invertible, this equation has a solution. Hence, the gauge \p{dD-g-vec} can indeed be imposed.

In terms of component vectors, the gauge condition \p{dD-g-vec} is represented by the system of
the following  $s{-}t{-}1$ gauge conditions:
\be\lb{dD-g-pm}
\begin{array}{rcl}
&& l|\varphi_{s}\rangle+A(0)|\varphi_{s-1}\rangle- l^+|\chi_{s-2}\rangle \ = \ 0 \,,\\[6pt]
&& l|\varphi_{s-1}\rangle+2A(1)|\varphi_{s-2}\rangle- l^+|\chi_{s-3}\rangle
-  A(0)|\chi_{s-2}\rangle \ = \ 0 \,,\\[6pt]
&& \ldots \ldots \\[6pt]
&& l|\varphi_{t+1}\rangle- l^+|\chi_{t-1}\rangle - A(s-t-2)|\chi_{t}\rangle \ = \ 0 \,.
\end{array}
\ee
As it is seen from \p{1-solut}, the condition \p{dD-g-vec} is equivalent to the condition
$|{\mathcal{Z}}_{s-1}\rangle=0$. This condition should preserve
the transformation law \p{GTAdS0a}. This means that the vector gauge parameter $|\Lambda_{s-1}\rangle$ should
be confined by the condition $L_0|\Lambda_{s-1}\rangle=0.$ The last condition yields the following conditions for
the component vectors $|\lambda_k\rangle$:
\be\lb{c-dD-la-pm}
\Bigl[l_0+A_0(s-k-1)\Bigr]|\lambda_{k}\rangle=0\,.
\ee

Gauge transformations of the other vectors $|\Phi_s\rangle$ and $|{\mathcal{X}}_{s-2}\rangle$ \p{GTAdS0}) with the confined parameter $|\Lambda_{s-1}\rangle$ for the component vectors $|\varphi_{s}\rangle$ and $|\varphi_{k}\rangle$ for t<k<s take the form
\begin{eqnarray}
\delta|\varphi_{s}\rangle &=& l^+|\lambda_{s-1}\rangle \,,
\lb{dFk-pm-s}
\\ [6pt]
\delta|\varphi_{k}\rangle &=&
l^+|\lambda_{k-1}\rangle+  A(s{-}k{-}1) \;|\lambda_{k}\rangle\,,
\qquad
t<k<s\,,
\label{dFk-pm}
\end{eqnarray}
where $|\lambda_{k}\rangle$ are given in \eqref{gauge-par-ex-pm} and satisfied to \p{c-dD-la-pm} and $ A$ are given in \eqref{sol-As}.
From \p{dFk-pm} it directly follows that we can remove the St\"{u}ckelberg fields $|\varphi_{k}\rangle$,
$t<k<s$ using their gauge transformations \eqref{dFk-pm}.

It works in the following way. Consider the last gauge transformation for $k=t+1$:
$\delta|\varphi_{t+1}\rangle=
l^+|\lambda_{t}\rangle+  A(s-t-2) \;|\lambda_{t+1}\rangle$ in  \eqref{dFk-pm} and impose
the gauge fixing condition $|\varphi_{t+1}\rangle=0$. The residual gauge transformations that preserve this gauge fixing condition must be subject to the condition
\be\lb{gf-la1}
|\lambda_{t+1}\rangle={}-\Bigl[ A(s-t-2)\Bigr]^{-1} l^+|\lambda_{t}\rangle
\ee
on the gauge parameters.
After that, using gauge transformations
$\delta|\varphi_{t+2}\rangle=
l^+|\lambda_{t+1}\rangle+  A(s-t-3) \;|\lambda_{t+2}\rangle$ from  \eqref{dFk-pm}, we can impose
the gauge fixing condition $|\varphi_{t+2}\rangle=0$. The residual gauge transformations that preserve this gauge fixing condition must be subject to the condition
\be\lb{gf-la2}
|\lambda_{t+2}\rangle={}-\Bigl[ A(s-t-3)\Bigr]^{-1} l^+|\lambda_{t+1}\rangle
= \Bigl[ A(s-t-2) A(s-t-3)\Bigr]^{-1} (l^+)^2|\lambda_{t}\rangle
\ee
on the gauge parameters, where expression \p{gf-la1} was used.
Continuing this procedure, we eliminate all the St\"{u}ckelberg fields
\be\lb{gf-varphi}
|\varphi_{k}\rangle=0\,,\qquad (t+1)\leq k\leq (s-1)
\ee
by imposing gauge fixing for all local transformations \p{dFk-pm}.
In these gauge fixings, the following relations for the gauge transformation parameters hold:
\be\lb{gf-la2}
|\lambda_{k}\rangle={}(-1)^{k-t}\Bigl[\prod\limits_{k=t+1}^{k}
 A(s-k-1)\Bigr]^{-1}\, (l^+)^{k-t}|\lambda_{t}\rangle
\,,\qquad (t+1)\leq k\leq (s-1)\,.
\ee
The remaining independent gauge parameter is $|\lambda_{t}\rangle$.

The gauge transformation \p{dFk-pm-s} of the remaining field
$|\varphi_{s}\rangle$ from the set
$|\Phi_s\rangle=\Bigl(|\varphi_{t+1}\rangle, \cdots , |\varphi_{s-1}\rangle,|\varphi_{s}\rangle\Bigr)$
\p{dFk-pm-s} has the form
\be
\delta|\varphi_{s}\rangle=
{}(-1)^{s-t-1}\Bigl[\prod\limits_{k=t+1}^{s-1}
 A(s-k-1)\Bigr]^{-1}\, (l^+)^{s-t}|\lambda_{t}\rangle \,.
\label{dFk-pm-s-1}
\ee
After rescaling $\displaystyle |\lambda_{t}\rangle\, \to\,
{}(-1)^{s-t-1}\prod\limits_{k=t+1}^{s-1}
 A(s-k-1)\, |\lambda_{t}\rangle$, transformations \p{dFk-pm-s-1}
are exactly the same as the transformations \p{tr-vect}:
\be
\delta|\varphi_{s}\rangle \ =
{}\  (l^+)^{s-t}|\lambda_{t}\rangle \,.
\label{dFk-pm-s-2}
\ee
It is important that the structure $\delta|\varphi_{s}\rangle\, \sim
\, (l^+)^{s-t}|\lambda_{t}\rangle$
arose due to the presence of the $(s-t-1)$
St\"{u}ckelberg fields
$|\varphi_{t+1}\rangle, |\varphi_{t+2}\rangle, \cdots , |\varphi_{s-2}\rangle, |\varphi_{s-1}\rangle$.

Furthermore, we have obtained a very important consequence for the transformation parameter $|\lambda_{t}\rangle$.
Taking into account the relation $A_0(k)=2\varkappa \, k(2s-k)$  \p{A0} and the condition \p{c-dD-la-pm}, one gets that the parameter
 $|\lambda_{t}\rangle$ is confined by the condition:
\be\lb{c-dD-la-k}
\Bigl[l_0 + 2\varkappa \, (s-t-1)(s+t+1)\Bigr]|\lambda_{t}\rangle=0\,.
\ee
This condition completely coincides with the first equation in \p{condition-vect} in the framework of the
mass-shell Fock space formulation.
But now it has been obtained directly within the framework of the BRST field description.

After elimination of the St\"{u}ckelberg fields
$|\varphi_{t+1}\rangle, \cdots , |\varphi_{s-1}\rangle$ by the gauge fixing conditions \p{gf-varphi},
the equations of motion \p{eqs-s} and \p{eqs-2s}
take the form:
\begin{eqnarray}
&&
\Bigl(l_0 -2\varkappa(2s+1) \Bigr)|\varphi_s\rangle \ = \ 0 \,,
\label{eqs-s-2}
\\ [5pt]
&&{}
\Bigl(l_0 + 2\varkappa \, (s-k-1)(s+k+1) \Bigr)|\chi_{k}\rangle
\ =\ 0\,,\qquad k=(t-1)\div(s-2)\,,
\label{eqs-2s-2}
\end{eqnarray}
where the relation $A_0(s-k-2)=2\varkappa \, (s-k-2)(s+k+2)$ has been used.

Applying the conditions \p{gf-varphi} to the  de Donder-like gauge conditions \p{dD-g-pm}, we obtain
\be\lb{dD-g-pm-1}
\begin{array}{rcl}
l|\varphi_{s}\rangle- l^+|\chi_{s-2}\rangle &=& 0 \,,\\[6pt]
l^+|\chi_{s-3}\rangle
+ A(0)|\chi_{s-2}\rangle &=& 0 \,,\\[6pt]
\ldots\ldots\ldots\ldots  & &   \\[6pt]
l^+|\chi_{t}\rangle
+ A(s-t-3)|\chi_{t+1}\rangle &=& 0 \,,\\[6pt]
l^+|\chi_{t-1}\rangle + A(s-t-2)|\chi_{t}\rangle &=& 0 \,.
\end{array}
\ee

Now we will show that all the fields $|{\mathcal{X}}_{s-2}\rangle=\Bigl(|\chi_{t-1}\rangle, \cdots , |\chi_{s-2}\rangle\Bigr)$ can be completely eliminated.  According to the second relation in \p{GTAdS0}, the gauge transformations of these fields have the form:
\begin{eqnarray}
\delta|\chi_{k}\rangle &=&
l |\lambda_{k+1}\rangle+ (s-k-1) A(s{-}k{-}2) \;|\lambda_{k}\rangle\,,
\qquad
t\leq k\leq (s-2) \,,
\lb{chi-pm-s}
\\ [6pt]
\delta|\chi_{t-1}\rangle &=&
l |\lambda_{t}\rangle\,.
\label{chi-pm}
\end{eqnarray}
where $|\lambda_{k}\rangle$ are given in \eqref{gf-la2} and the set of quantities $A$ is given in \eqref{sol-As}.
We see that the field $|\chi_{t-1}\rangle$ can be gauged away by the transverse part in $|\lambda_{t}\rangle$.
That is, we obtain
\be
|\chi_{t-1}\rangle\ =\ 0
\label{chi-t1}
\ee
after fixing this gauge.
Then, as follows from \p{chi-pm},
there is a residual gauge symmetry with the parameter $|\lambda_{t}\rangle$ that obeys \p{c-dD-la-k} and the condition
\be
l |\lambda_{t}\rangle\ =\ 0\,.
\label{la-long}
\ee
In addition, if condition \p{chi-t1} is fulfilled, then we directly obtain from relations \p{dD-g-pm-1} that
all the fields $|\chi_{k}\rangle$ vanish
\be
|\chi_{k}\rangle\ =\ 0\,,\qquad
(t-1)\leq k\leq (s-2)\,,
\label{chi-tk}
\ee
and the field $|\varphi_{s}\rangle$ is subject to the condition
\be
l|\varphi_{s}\rangle\ =\ 0\,.
\label{phi-s-long}
\ee
As a result, we have obtained the second of the relations \p{condition-vect} defining the mass-shell of the partially massless theory.

Thus, the mass shell of the model under consideration
is described by the field $|\varphi_s\rangle$,
which is subject to equations \p{eqs-s-2} and \p{phi-s-long}.
These equations coincide
with the conditions \p{condition-vect} defining the irreducible partially massless representation in $dS$ space.
Gauge transformations \p{dFk-pm-s-2}  coincide with gauge transformations  \p{tr-vect},
the parameter of which is limited by conditions \p{c-dD-la-k} and \p{la-long}. The formulation of the
BRST Lagrangian for the partially massless higher-spin field theory is complete.

Note that in the formulation under consideration in terms of the Fock space,
the expansion of the obtained spectrum into helicity states is quite clear.
Indeed, let us write the expansions
\be\lb{st-phi-tr-s}
|\varphi_{s}\rangle \ =\ \underbrace{|\phi_{s}\rangle
+\ldots +(l^+)^{s-t-1}|\phi_{t+1}\rangle}_{\mbox{physical states}}
+\underbrace{(l^+)^{s-t}|\phi_{t}\rangle +\ldots  + (l^+)^s|\phi_{0}\rangle}_{\mbox{gauge states}}\,,
\ee
\be\lb{st-la-tr-s}
|\lambda_{t}\rangle \ =\ |\xi_{t}\rangle+l^+|\xi_{t-1}\rangle
+\ldots + (l^+)^{t-1}|\xi_{1}\rangle + (l^+)^t|\xi_{0}\rangle\,,
\ee
where all vectors in these expansions are transversal:
\be\lb{st-tr-s}
l\,|\phi_{k}\rangle = 0\,,\qquad l\,|\xi_{k}\rangle = 0\,.
\ee
Then, by using gauge transformations \p{dFk-pm-s-2}, namely
$\delta|\varphi_{s}\rangle =  (l^+)^{s-t}|\lambda_{t}\rangle$,
we remove the fields $|\phi_{t}\rangle,\ldots,|\phi_{0}\rangle$.
As a result, we are left with the fields $|\phi_{s}\rangle,|\phi_{s-1}\rangle,
\ldots, |\phi_{t+2}\rangle, |\phi_{t+1}\rangle$, which describe the states with helicities
$\pm s,\pm(s-1),\ldots,\pm(t+2),\pm(t+1)$.

\setcounter{equation}{0}
\section{Component Lagrangian}

As we see, all relations are much simpler when we deal with the Fock space vectors. Moreover,  conventional field
theory is, in a certain sense, the same BRST theory just rewritten in ``component form''. Since the field
formulation is often more commonly used, we will demonstrate here the transition from the formulation in terms
of the Fock space vectors to the formulation in terms of multi-spinor fields.

For transition to the component field formulation, we use only the following obvious relations:
\be\lb{norm-b-k}
\langle0|(b)^k(b^+)^n|0\rangle={}\delta_{kn}k!\,,
\ee
\be\lb{norm-ac-k}
\langle0|a_{\alpha_1}\ldots a_{\alpha_k}\bar a^{\dot\alpha_1}\ldots \bar a^{\dot\alpha_k}
c^{\beta_1}\ldots c^{\beta_n}\bar c_{\dot\beta_1}\ldots \bar c_{\dot\beta_n}|0\rangle={}
\delta_{kn}(k!)^2\delta_{(\alpha_1}^{\beta_1}\ldots\delta_{\alpha_k)}^{\beta_k}
\delta^{(\dot\alpha_1}_{\dot\beta_1}\ldots\delta^{\dot\alpha_k)}_{\dot\beta_k}\,.
\ee

Taking into account expansions \p{GFState-pm}, \p{GFState-pm-1} and using \p{norm-b-k}, \p{norm-ac-k}, the Lagrangian \p{Lagr-vector-1} takes the following form:
\begin{eqnarray}
\nonumber
\mathcal{L}_{s,t}
&=&{}\sum\limits_{k=\,t+1}^{s}(s-k)!\,\bar\varphi^{\alpha(k)}_{\dot\beta(k)}
\Bigg( \frac{2k+1}{k}\, \Box - m_t^2 +\varkappa\left(2s^2+k \right)
+\frac{(s-k)A^2(s-k-1)}{2k+2} \Bigg)\varphi_{\alpha(k)}^{\dot\beta(k)}
\\ [5pt]
\nonumber {}
&&{}-\sum\limits_{k=\,t+1}^{s}\frac{(s-k)!(k+1)^2}{2k}\,\bar\varphi^{(\alpha(k)}_{(\dot\beta(k)}
\nabla^{\alpha)}_{\dot\beta)}\nabla_{(\alpha}^{(\dot\beta}\varphi_{\alpha(k))}^{\dot\beta(k))}
\\ [5pt]
&&\nonumber {}
+\sum\limits_{k=\,t+1}^{s}\frac{(s-k+1)!}{2}\,A(s-k)\Bigg(
\bar\varphi^{\alpha(k)}_{\dot\beta(k)}
\nabla_{(\alpha}^{(\dot\beta}\varphi_{\alpha(k-1))}^{\dot\beta(k-1))}
-\bar\varphi^{(\alpha(k-1)}_{(\dot\beta(k-1)}
\nabla^{\alpha)}_{\dot\beta)} \varphi_{\alpha(k)}^{\dot\beta(k)}\Bigg)
\\ [5pt]
&&\nonumber {}
-\sum\limits_{k=\,t-1}^{s-2}(s-k-2)!\,\bar\chi^{\alpha(k)}_{\dot\beta(k)}
\Bigg( \Box - m_t^2 +\varkappa\left[2s^2-(k+1)^2-1 \right]
\\ [5pt]
&&\nonumber {}
\hspace{8cm}
+\frac{(s-k)A^2(s-k-1)}{2k+2} \Bigg)\chi_{\alpha(k)}^{\dot\beta(k)}
\\ [5pt]
\nonumber {}
&&{}-\sum\limits_{k=\,t-1}^{s-2}\frac{(s-k-2)!(k+1)^2}{2k}\,\bar\chi^{(\alpha(k)}_{(\dot\beta(k)}
\nabla^{\alpha)}_{\dot\beta)}\nabla_{(\alpha}^{(\dot\beta}\chi_{\alpha(k))}^{\dot\beta(k))}
\\ [5pt]
&&{}
+\sum\limits_{k=\,t-1}^{s-2}\frac{k(s-k-1)!}{2k+2}\, A(s-k-2)\Bigg(\bar\chi^{\alpha(k)}_{\dot\beta(k)}
\nabla_{(\alpha}^{(\dot\beta}\chi_{\alpha(k-1))}^{\dot\beta(k-1))}
- \bar\chi^{(\alpha(k-1)}_{(\dot\beta(k-1)}
\nabla^{\alpha)}_{\dot\beta)} \chi_{\alpha(k)}^{\dot\beta(k)} \Bigg)
\nonumber
\\ [5pt]
&&{}
-\sum\limits_{k=\,t+1}^{s}\frac{(k-1)(s-k)!}{2}\,\Bigg(
\bar\varphi^{\alpha(k)}_{\dot\beta(k)}
\nabla_{(\alpha_1}^{(\dot\beta_1}\nabla_{\alpha_2}^{\dot\beta_2}\chi_{\alpha(k-2))}^{\dot\beta(k-2))}
+\bar\chi^{(\alpha(k-2)}_{(\dot\beta(k-2)}
\nabla^{\alpha_1}_{\dot\beta_1}\nabla^{\alpha_2)}_{\dot\beta_2)} \varphi_{\alpha(k)}^{\dot\beta(k)} \Bigg)
\nonumber
\\ [5pt]
&&{}
-\sum\limits_{k=\,t+1}^{s-2}\frac{(s-k)!}{2k+2}\,A(s-k-1)A(s-k-2)\Bigg(
\bar\chi^{\alpha(k)}_{\dot\beta(k)}\varphi_{\alpha(k)}^{\dot\beta(k)}
+\bar\varphi^{\alpha(k)}_{\dot\beta(k)}\chi_{\alpha(k)}^{\dot\beta(k)} \Bigg)
\nonumber
\\ [5pt]
&&{}
+\sum\limits_{k=\,t+1}^{s-1}\frac{(2k+1)(s-k)!}{2(k+1)}\,A(s-k-1)\Bigg(
\bar\chi^{(\alpha(k-1)}_{(\dot\beta(k-1)}
\nabla^{\alpha)}_{\dot\beta)} \varphi_{\alpha(k)}^{\dot\beta(k)}
+ \bar\varphi^{\alpha(k)}_{\dot\beta(k)}
\nabla_{(\alpha}^{(\dot\beta}\chi_{\alpha(k-1))}^{\dot\beta(k-1))} \Bigg) \,,
\label{Lagr-vector-comp}
\end{eqnarray}
where the coefficients $A(p)$ at the fixed value $p$ are defined in \p{Ak-k}.

The component field Lagrangian \p{Lagr-vector-comp} is invariant with respect to gauge transformations
\begin{eqnarray}
\delta \varphi_{\alpha(s)}^{\dot\beta(s)} &=& s\nabla_{(\alpha}^{(\dot\beta}\lambda_{\alpha(s-1))}^{\dot\beta(s-1))} \,,
\lb{dFk-comp-s}
\\ [6pt]
\delta \varphi_{\alpha(k)}^{\dot\beta(k)} &=&
k\nabla_{(\alpha}^{(\dot\beta}\lambda_{\alpha(k-1))}^{\dot\beta(k-1))} +  A(s{-}k{-}1) \;\lambda_{\alpha(k)}^{\dot\beta(k)}\,,
\qquad
t<k<s\,,
\label{dFk-comp}
\end{eqnarray}
\be\lb{var-chi}
\delta \chi_{\alpha(k)}^{\dot\beta(k)} ={}-\frac{1}{k+1}\,
\nabla_{\dot\gamma}^{\gamma}\lambda_{(\gamma\alpha(k+1))}^{(\dot\gamma\dot\beta(k+1))} + (s{-}k{-}1)A(s{-}k{-}2) \;\lambda_{\alpha(k)}^{\dot\beta(k)}\,,
\qquad
(t-1)\leq k\leq (s-2)\,,
\ee
which are the transformations \p{GTAdS0} (see also \p{dFk-pm-s} and \p{dFk-pm}) in component form.
Note that due to \p{Ak0}, some terms in the variations \p{dFk-comp} and \p{var-chi} are zero.
Let us emphasize that the component Lagrangian \p{Lagr-vector-comp}, although it looks rather cumbersome,
is nothing more than another form of the Lagrangian \p{actionQ}.

\setcounter{equation}0
\section{Summary}

In this paper, we have presented the self-consistent BRST Lagrangian formulation for partially massless
bosonic fields in four-dimensional space. All details of this formulation, including the
the correct de Sitter vacuum, were derived in a unified manner from the requirement that the BRST charge
be Hermitian and nilpotent.

We emphasize the main features and results of the approach under consideration.

\begin{itemize}
\item
The conditions defining the mass-shell of the partially massless fields in (A)dS are formulated in
terms of two-component spin-tensor fields. In four dimensions, such a formulation is much more
convenient than one in terms of conventional tensor fields since the tracelessness condition is
automatically fulfilled.
\item
The mass shell conditions are reformulated in terms of the Fock space vectors \p{GFState}.
The operators \p{op-0}, \p{op-1}, \p{op-t1}, acting in the Fock space are introduced which makes it possible to rewrite
the mass shell conditions \p{condition1} in the form of constraints \p{condition-vect} on
the Fock space vectors. The gauge transformations \p{tr-sp} and conditions on gauge parameters are also
rewritten in terms of the Fock space as the constraints \p{tr-vect} and \p{condition-vect-la}.
\item
In the partially massless case, the mass parameter $m_t$ \p{m-t} is nonzero. In this case, it turned
out that the basic constraints \p{op-0}, \p{op-1}, \p{op-t1} correspond to the system with
second-class constraints, and to construct the BRST charge we have developed the conversion procedure to
obtain the first-class constraints system. It is shown that the hermitianity and nilpotency of the BRST charge
lead to restrictions on the parameters of the theory that are satisfied only for dS space. AdS space is excluded.
Essentially, the requirements of hermitianity and nilpotency of the BRST charge turn out to be equivalent to the conditions of the
unitary representation for partially massless fields in dS space.
\item
The hermitianity of the BRST charge in the theory under consideration requires that the vectors used
be restricted to expansions \p{GFState-pm} and \p{GFState-pm-1}. Moreover, the presence of $(s-t-1)$ St\"{u}ckelberg
fields in expansion \p{GFState-pm} leads, after their
exclusion, to gauge transformations \p{dFk-pm-s-2} of physical fields of the $(s-t)$ order in
derivatives.
\item
The final Lagrangian \p{Lagr-vector-1}, obtained within the BRST formalism, describes on the mass-shell only
the states with helicities $\pm s,\pm(s-1),\ldots,\pm(t+2),\pm(t+1)$. No other states remain in
the theory after eliminating all auxiliary and gauge degrees of freedom. The resulting Lagrangian is rewritten in terms of conventional spin-tensor fields in the form \p{Lagr-vector-comp} and is a generalization of the Fronsdal Lagrangian to the partially massless case.
\end{itemize}

The Lagrangian formulation of partially massless fields developed here can be extended in different directions. We point out here some of them.
\begin{itemize}
\item
Lagrangian formulation for fermionic partially massless fields in dS${}_4$ space. Like in the bosonic case, the BRST approach can allow  constructing the Lagrangian formulation only in terms of partially massless fields.
\item
Construction of cubic vertices for interacting partially massless fields among themselves as well as with conventional massless and massive higher spin fields.
\end{itemize}
We plan to study all these items in the forthcoming papers.

\section*{Acknowledgments}
The authors are grateful to K.B. Alkalaev, A.O. Barvinsky, M.A. Vasiliev, B.L. Voronov and Yu.M. Zinoviev for discussion and useful comments.

\end{document}